%% file: JSA_Soccer_v0.1.tex
\documentclass[Royal,sageh,times]{sagej}

\usepackage{moreverb,url}
\usepackage{placeins}
\usepackage[colorlinks,bookmarksopen,bookmarksnumbered,citecolor=red,urlcolor=red]{hyperref}

\newcommand\BibTeX{{\rmfamily B\kern-.05em \textsc{i\kern-.025em b}\kern-.08em
T\kern-.1667em\lower.7ex\hbox{E}\kern-.125emX}}

\def\volumeyear{2026}
\global\long\def\bX{\mathbf{X}}%

\begin{document}


\title{Framing Causal Questions in Sports Analytics: A Tutorial on Estimand Choice Illustrated Through Crossing in Soccer}

\author{Shomoita Alam\affilnum{1}, Erica E. M. Moodie\affilnum{1}, Lucas Y. Wu\affilnum{2} and Tim B. Swartz\affilnum{3}}

\affiliation{\affilnum{1}McGill University, Department of Epidemiology, Biostatistics and Occupational Health, Montreal, Canada\\
\affilnum{2}Teamworks, Burnaby, Canada\\
\affilnum{3}Simon Fraser University, Department of Statistics and Actuarial Science, Burnaby, Canada}

\corrauth{Shomoita Alam, McGill University, Department of Epidemiology, Biostatistics and Occupational Health, Montreal, Canada}
\email{shomoita.alam@mail.mcgill.ca}

\begin{abstract}
Causal inference has become an accepted analytic framework in sports analytics, where experimentation is rarely feasible. A key consideration is the choice of estimand, specifically, whether to target the Average Treatment Effect (ATE), which reflects the effect of an action across the entire population, or the Average Treatment Effect on the Treated (ATT), which reflects the effect among those who actually took the action. Using data from nearly all 240 matches of the 2019 Chinese Super League season, we apply propensity score matching to estimate the causal effect of crossing on shot creation in soccer. The ATE and ATT are nearly identical (0.033 and 0.035 respectively), a result we attribute to substantial overlap in propensity score distributions between plays where a cross was and was not attempted. To illustrate when these estimands diverge, we construct two simulation scenarios with known ground truth: one reproducing the high-overlap structure of the real data, where ATE and ATT coincide, and one engineered to exhibit severe confounding and low overlap, where they diverge substantially. While empirical findings are specific to the 2019 Chinese Super League season, the case study and simulations provide a principled guide to estimand choice in causal analyses of sports data. 
\end{abstract}


\maketitle

\section{Introduction \label{sec:Intro}}

Over the past quarter century, statistical causal inference has made significant 
advancements, leading to its widespread adoption across diverse disciplines, 
including healthcare, economics, and education. These methods offer a framework 
to uncover relationships that go beyond associations, enabling researchers to 
address causal questions directly. However, many studies proceed with selecting 
estimation techniques without first clearly articulating the causal question or 
specifying the target of estimation \citep{goetghebeur2020formulating}. This 
lack of clarity can undermine the reliability, interpretability, or comparability 
of conclusions.

Existing overview papers on causal inference often focus on the properties of 
specific methods but fail to address broader issues, such as whether the chosen 
technique aligns with the study's goals or whether the necessary assumptions for 
causal validity are satisfied. This creates challenges for researchers navigating 
an expanding methodological landscape while attempting to apply these tools 
effectively to their questions. One such area is sports analytics, where causal 
inference is increasingly recognized as a powerful but underused tool for 
data-driven decision-making. See \cite{lopez2020bigger, wu2021contextual, 
nakahara2023pitching, dona2023causal, epasinghege2024causal, gibbs2022causal}; and \citet{yam2019lost} for some rare examples of causal analyses in sport. 
Sports provide a structured and well-defined setting for causal analysis, with 
clear objectives, fixed timeframes, and established rules. These characteristics 
facilitate the identification of confounding variables, a key challenge in causal 
inference, making sports a more tractable domain than many complex scientific 
fields. This clarity enhances the robustness and interpretability of causal 
estimates, reinforcing the value of causal methods in sports analytics.

In this article, we provide a tutorial for sports researchers in the form of a 
case study and simulations, focusing on framing causal questions and answering them using causal 
inference tools. We illustrate this through an analysis investigating whether crossing is an effective strategy for creating goal-scoring opportunities (shots) in soccer. In soccer or association football, a ``cross'' is defined as 
a long pass delivered from a wide area of the field toward a central area near 
the opponent's goal. Crossing has traditionally been regarded as a key offensive 
strategy, based on the assumption of a positive correlation between the frequency 
of crosses and the number of goals scored. However, recent studies have 
challenged this belief, presenting evidence of a negative association between 
open-play crosses and goal-scoring effectiveness 
\citep{vecer2014crossing, sarkar2018paradox}.

The dataset that we shall employ to motivate this tutorial includes detailed 
tracking data on crossing opportunities and player movements in potential 
crossing zones from nearly all 240 matches of the 2019 Chinese Super League 
season, spanning all teams in the league. It captures 50,550 crossing 
opportunities, of which 5,614 involved an actual cross and 44,936 did not. We 
take as our outcome the binary indicator of whether a shot resulted from the 
play. That is, we will focus on whether a shot was taken rather than whether 
the shot created a goal-scoring opportunity. Additionally, several potential 
confounders - that is, variables that are believed to affect both the decision 
to cross or not and the probability of a shot being taken - are included which 
were derived from spatiotemporal tracking data. A subtlety that is often 
overlooked in applied causal analyses is the distinction between the Average 
Treatment Effect (ATE) and the Average Treatment Effect on the Treated (ATT). In our case study, we find 
that the ATE and ATT are nearly identical, a result we attribute to the 
substantial overlap in propensity score distributions between plays where a cross 
was and was not attempted. To more clearly illustrate the conditions under which 
these estimands diverge, we complement the case study with two simulation 
scenarios constructed to represent, respectively, a setting where ATE and ATT 
coincide and one where they do not.

Utilizing the case study and simulation results, this paper provides a step-by-step guide to formulating the research question in 
the causal framework, emphasizing the importance of specifying precise causal 
questions, defining estimation targets, and evaluating assumptions before 
selecting an analysis method. By bridging the gap between theory and practice, 
this tutorial aims to equip practitioners in sports analytics with the tools to 
make rigorous and informed methodological choices tailored to their specific 
research goals.

The remainder of this paper is organized as follows. 
Section~\ref{sec:framework} outlines the components of the causal inference 
framework, including the formulation of the intervention of interest, outcome, 
and potential outcomes, as well as the assumptions required for causal validity. 
Section~\ref{analysis_methods} describes propensity score matching and its 
implementation for estimating the ATE and ATT. Section~\ref{sec:simulation} 
introduces the simulation study, demonstrating through two contrasting scenarios 
when and why the ATE and ATT diverge. Section~\ref{sec:casestudy} presents the 
empirical case study, applying the methods of 
Sections~\ref{analysis_methods} and~\ref{sec:simulation} to data from the 2019 
Chinese Super League season. Finally, Section~\ref{sec:discussion} discusses the 
broader methodological and practical insights, highlighting the strengths and 
limitations of the approach and providing recommendations for future work in 
sports analytics.

\section{Components of the potential outcomes framework for causal 
inference \label{sec:framework}}

The dominant statistical framework for causal inference over the last three 
decades is the potential outcomes framework, also known as the Rubin Causal 
Model \citep{neyman1923application, rubin1974estimating}. This framework is 
built on the concept that each individual in a study has \textit{potential 
outcomes} that correspond to each of the possible treatments of interest: in 
the case of a binary treatment, with would be the outcome that would be observed 
if they received the treatment, and the outcome that would be observed if they 
did not receive the treatment. The significance to this approach is that it 
forces the researcher to clearly specify the ``treatment'' alternatives (e.g., 
decisions, actions) under consideration, and to describe the effects as contrasts 
of outcomes attributable \textit{only} to differences in these alternatives-all 
other factors are unchanged in defining this contrast. Any observed effect should 
then be attributable only to the differences in treatment alternatives, as there 
are no other factors which vary.

For the soccer crossing research, causal questions can be formulated in two 
fundamentally distinct ways: (a) What is the effect of crossing if applied to 
all plays? (b) What is the effect of crossing for plays where it was in fact 
used? The first question addresses the effectiveness of crossing to create shot 
opportunities across the entire population of plays or crossing opportunities, 
while the second question focuses on the effectiveness specifically in situations 
where crossing has already been attempted. To consider a medical analogy which 
may help to clarify the distinction, consider a non-randomized study of a 
smoking cessation treatment. The first question, the ATE, seeks to understand 
the impact of the treatment across the entire population of smokers. The second, 
the ATT, instead seeks to estimate the impact only among those smokers who 
actually received the treatment. The population of all smokers and the population 
of smokers who sought treatment may not be identical - for instance, those who 
sought treatment may be more motivated to quit, and this motivation may itself 
influence the treatment's effectiveness, so that the average effect among those 
treated differs from the average effect across the whole population.

To empirically evaluate these questions, it is necessary to define the key 
estimands. In the following sections, we discuss the fundamental components 
needed to formalize the causal question in the crossing problem. Note that much 
of the causal inference language is described in terms of medical interventions; 
we clarify how this translates to sporting analyses.

\begin{enumerate}
    \item {\textbf{Treatment: }} A treatment refers to an action or 
    intervention under study. In this study, the treatment variable is whether 
    the player makes an open-play cross or not. This is represented by the 
    binary random variable $Z$, where $Z=1$ if a cross is made, and $Z=0$ 
    otherwise.

    \item {\textbf{Outcome:}} The outcome is defined as whether a crossing 
    opportunity results in a shot. We denote this outcome variable by $Y$, taking the value 1 if a shot was made and 0 otherwise. Note that, a shot is any attempt directed towards the opponent's goal, including headers, regardless of whether it was on target, but excluding efforts ruled offside.

    \item {\textbf{Population of interest:}} The population of interest can 
    vary depending on the causal question. In the context of our motivating 
    analysis, the population refers not to the players but rather the plays 
    themselves, or the crossing opportunities. If we are interested in the 
    effect of crossing when applied to all plays, the population of interest is 
    the entire dataset of crossing opportunities. Conversely, if the interest 
    lies in the effect of crossing in plays where crossing is actually attempted, 
    the population of interest consists of those specific instances where 
    crossing was used. Generally, we require that each unit or member of the 
    population of interest (in this case, each play), has a non-zero 
    probability of either crossing or not crossing, a requirement known as the \textit{positivity 
    assumption} \citep{petersen2012diagnosing}. This requires, for each 
    combination of situations in which a play occurred (where a ``situation'' 
    is defined by a combination of factors such as the distance between sender 
    and nearest teammate or the offensive-defensive ratio in box; see item 5, 
    Confounders, below), that there are some plays in which a cross was taken 
    and also some plays under identical situation where a cross was not.

    \item {\textbf{Potential outcomes:}} For the observed outcome $Y$, we 
    define $Y = Y(1)$ if $Z=1$, and $Y=Y(0)$ if $Z=0$. Here, $Y(0)$ and 
    $Y(1)$ are the potential outcomes under no crossing and crossing, 
    respectively. For units where $Z=1$, we observe $Y(1)$, and $Y(0)$ is an 
    unobserved counterfactual. Similarly, for units with $Z=0$, we only observe 
    $Y(0)$ while $Y(1)$ is unobserved. Under the potential outcomes framework, 
    the problem of causal inference becomes one of estimating quantities related 
    to unobserved counterfactuals using only observed data.

    We rely on certain assumptions to define the potential outcomes, such as 
    \textit{no interference}. This assumption requires that, for a given play, 
    the outcome may be influenced by the act of crossing (or not) and that the 
    outcome is not affected by the decision to cross (or not) \textit{at any 
    other play}. It is possible that this assumption may not hold in the 
    crossing problem in our motivating example due to tactical 
    interdependencies, player interactions, player psychology, and sequential 
    play dynamics. For instance, teams may adjust their defensive or offensive 
    strategies in response to earlier plays, or a cross may pull defenders out 
    of position, influencing subsequent outcomes. We will nevertheless proceed under the assumption of no interference, which we consider reasonable, focusing on the immediate effect 
    of a cross (i.e., treating each crossing opportunity as a conditionally 
    independent unit) and ensuring the effect of other variables such as 
    proximity of defenders is adequately taken into account by our modeling 
    approach.

    The definition of potential outcomes given above follow directly from the 
    axiom of \textit{causal consistency}, which states that the observed outcome 
    corresponds to the outcome expected under the action under consideration. In 
    formal terms, this implies $Y = Z \cdot Y(1) + (1 - Z) \cdot Y(0)$, where 
    $Z$ indicates the action of crossing or not. For this axiom to be taken as 
    true, two key conditions must hold: (1) \textit{treatment variation 
    irrelevance}, meaning all ``crosses'' are treated as equivalent forms of the 
    same action and similarly for instances of not crossing, and (2) \textit{no 
    interference}, explained above. In practice, we aim to ensure treatment 
    irrelevance holds by carefully defining both the levels of the treatment 
    (crossing or not) and the outcome (whether a shot was taken) such that any 
    differences in how a cross is executed do not meaningfully alter its 
    classification as a ``cross.'' Under these conditions, causal consistency 
    allows us to interpret the observed outcome as the potential outcome for the 
    relevant action scenario, thereby supporting valid comparisons of what the 
    effect of crossing versus not.

    \item {\textbf{Confounders:}} Let $\bX$ denote the vector of potential 
    confounders. A confounder is a variable that influences both the treatment 
    (action of crossing) and the outcome (shot), potentially creating a spurious 
    association that can bias causal effect estimates. In randomized controlled 
    trials (RCTs), treatments are randomly assigned, ensuring that there are no 
    confounders, by virtue of the treatment allocation mechanism being 
    (typically) a simple randomization that is independent of any covariates. 
    However, in the context of the crossing analysis in soccer - and indeed 
    virtually all other in-game actions, randomization is not feasible since the 
    decision to cross cannot be externally assigned during a match. Therefore, 
    the causal effect must be studied using an `observational' 
    (non-experimental) framework, incorporating methods to adjust for 
    confounders. Indeed, causal inference analyses all, fundamentally, aim to 
    perform an analysis that mimics an RCT, analytically creating 
    exchangeability between those instances where $Z=1$ and those where $Z=0$.

    In the crossing problem, potential confounders include the score 
    differential, the estimated time for the nearest defender to reach the ball 
    carrier, the distance of the ball carrier to the nearest sideline, the 
    distance of the nearest teammate to the opponent's goal line, the distance 
    of the ball carrier to the opponent's goal line, the ratio of offensive to 
    defensive players in the box, the position of the crosser, and whether the 
    play occurred in the last ten minutes of a half. All distances are measured 
    in meters.

    \item {\textbf{Causal parameter or estimand:}} Population summary 
    measures can be estimated for different groups, such as the entire 
    population or the subpopulation of plays in which a cross did or did not 
    occur (in the more typical language of causal inference, this would 
    correspond to treated or untreated individuals). In the context of the 
    soccer data, we focus on the mean difference as the primary causal contrast 
    of interest, which in this case is equivalent to a difference in probability 
    since the outcome is binary.

    The \textit{Average Treatment Effect} is the estimand which addresses the 
    question: ``What would be the average number of shots if crossing was 
    attempted in all plays, versus the average number of shots if crossing was 
    not attempted in any play?'' It is defined as:
    \[
    ATE = \mathbb{E}[Y(1)] - \mathbb{E}[Y(0)].
    \]
    Note that the ATE is a marginal quantity, averaging across the possible 
    situations in which a crossing opportunity was defined.

    In contrast, the \textit{Average Treatment Effect on the Treated} addresses 
    the question: ``Among plays where a cross was observed to have been taken, 
    what was the difference in the average number of shots attributable to 
    crossing the ball relative to not crossing?'' It is defined as:
    \[
    ATT = \mathbb{E}[Y(1) \mid Z=1] - \mathbb{E}[Y(0) \mid Z=1].
    \]
    The ATT focuses on the effect of crossing for those plays where a cross was 
    actually taken, measuring the difference between the observed outcome and 
    the counterfactual outcome had they not crossed. The difference between the 
    ATE and the ATT is subtle but important. From a purely statistical 
    perspective, the difference lies in a marginal expectation as opposed to one 
    that is conditional on $Z=1$. As $Z$ is affected by confounders $\bX$, the 
    distribution of $\bX$ as compared to the distribution of $\bX|Z=1$ will 
    differ, and so the expectation of $Y(1)-Y(0)$ under the marginal and 
    conditional distributions will differ. From a subject-matter perspective, a 
    cross is much more likely to be taken, for example, when the space is 
    controlled by the sender (player taking the play and hence deciding whether 
    to cross) as well as when there is a higher offensive-defensive player ratio 
    in the box.

    Another related estimand is the \textit{Average Treatment Effect on the 
    Non-Treated} (ATNT), which examines the effect of crossing in plays where 
    in fact the players chose not to cross the ball. It addresses the question: 
    ``Among the plays where crossing was not attempted, what would be the 
    average number of shots on goal if crossing was attempted, compared to if 
    it was not?'' It is defined as:
    \[
    ATNT = \mathbb{E}[Y(1) \mid Z=0] - \mathbb{E}[Y(0) \mid Z=0].
    \]
    The ATNT evaluates the potential benefit of crossing for plays where 
    crossing was not attempted, comparing the hypothetical outcomes with the 
    observed ones. In the next section, we focus on the estimands ATE and ATT, 
    but the ATNT can also be analyzed using similar methods.\\

    Before proceeding to the analysis, it is worth consolidating the key assumptions that underpin the identification of these estimands from observational data. Together, the components outlined above, i.e., the treatment, outcome, population of interest, potential outcomes, and confounders, form the building blocks of the causal inference framework. However, moving from this framework to valid causal estimates requires four assumptions to hold: (1) \textit{no interference}, which requires that the outcome of one play is not affected by the crossing decision in any other play; (2) \textit{causal consistency}, which requires that the observed outcome corresponds to the potential outcome under the action actually taken; (3) the \textit{positivity assumption}, which requires that each play has a non-zero probability of both crossing and not crossing; and (4) \textit{no unmeasured confounding} (NUC), discussed further in Section~\ref{analysis_methods}, which requires that all relevant confounders have been measured and included in the analysis. While we proceed under all four assumptions, we note that no interference and NUC are the most difficult to verify in practice, as acknowledged in Section~\ref{sec:discussion}.

\end{enumerate}

\section{Analysis using propensity score matching: a causal inference 
approach \label{analysis_methods}}

\subsection{Confounding in observational data}

A central challenge in identifying causal effects from observational data is 
the presence of \emph{confounding}. In the context of crossing decisions in 
soccer, a confounder is any factor that influences both the likelihood of 
attempting a cross (the ``treatment'') and the likelihood of creating a shot 
(the outcome). Because crosses are not randomly assigned, players decide to 
cross (or not) based on situational factors such as the distance to the nearest 
defender or the ratio of offensive to defensive players in the box. If these 
factors also influence whether a shot is taken, failing to account for them can 
bias the estimated effect of crossing.

The extent of imbalance in a covariate between two treatment or action groups 
is often captured via the standardized absolute mean difference (SMD), a measure 
of the difference in mean or proportion across the two groups, scaled by a 
pooled measure of variability; values of 10\% or higher are considered 
indicative of poor covariate balance. In Section~\ref{sec:casestudy}, we 
illustrate the presence of such confounding in our soccer dataset, demonstrating 
how situational variables differ substantially between plays with and without a 
cross. We first describe the general matching strategy before applying it to the crossing data in Section~\ref{sec:casestudy}.

\subsection{Matching-based analytical strategy}

In this section, we discuss estimating the causal effects of crossing using a 
particular causal inference approach that relies on matching. Several alternative 
analytic approaches exist \citep{goetghebeur2020formulating} that can also be 
employed to have the same aim - to reduce differences (imbalance) in the 
distribution of confounders between the groups defined by the action of crossing 
(i.e., whether or not the ball was crossed). Many approaches rely on the 
\textit{no unmeasured confounding} (NUC) assumption, which states, in the 
context of crossing, that the action of crossing is independent of potential 
outcomes (i.e., whether or not there was a shot), given the confounders $\bX$. 
Formally, this assumption can be written as:
\[
\{Y(0), Y(1)\} \perp Z \mid \bX 
\]
where $Z$ represents the binary variable for the choice of action. Essentially, 
NUC implies that all relevant confounding variables $\bX$ have been measured 
and are available for analysis. By accounting for this set of confounders, we 
can estimate the causal effect by comparing observed outcomes between the plays 
where crossing was attempted with the outcome of the plays where crossing was 
not attempted but that are otherwise (nearly) identical with respect to $\bX$. 
That is, we are attempting to identify pairs of plays where $\bX$ holds and 
$Z=1$ for one play in the pair but $Z=0$ for the other pair such that the 
outcomes for these two plays can serve as the missing counterfactuals in the 
contrast $Y(1)-Y(0)$.

One common method of adjusting for confounding is \textit{propensity score 
matching}, which refers to the `propensity' to take the action of interest - 
in our case, the propensity for a cross to be taken during a play. This approach 
relies on the NUC assumption. In our analysis, we use propensity score matching 
to estimate both of the causal effects for the causal questions posed in the 
previous section. In the following subsections, we discuss the propensity score 
model, and propensity score matching to estimate different estimands of interest 
and covariate balance, highlighting the differences in the way the matching is 
carried out to accomplish estimation of the ATE and the ATT.

\subsection{The propensity score}

The propensity score is a function of covariates that reduces the vector $\bX$ 
into a single value, capturing all measured information that is relevant to the 
choice of actions \citep{rosenbaum1983central}. This score has the 
\textit{balancing property}, meaning that, conditional on the propensity score, 
the confounder distributions between the plays where crossing was or was not 
attempted are similar, rendering them effectively exchangeable. Typically, the 
propensity score is estimated using a parametric model, such as logistic 
regression, or a non-parametric approach like tree-based classification 
\citep{breiman2001random}, to predict the probability that the sender will cross 
the ball given the confounders. Being a probability, the propensity score lies 
in the range [0,1]. \citet{alam2019should} demonstrated that a simple 
parametric approach such as logistic regression often outperforms more complex 
models for propensity score estimation in terms of covariate balance. See also 
\citet{austin2011introduction} and \citet{stuart2010matching} for broader 
overviews of propensity score methods.

Regardless of the method used to estimate the propensity score, its adequacy 
must be assessed by evaluating the balance of confounders across the different 
choices of actions \citep{tan2006distributional} following matching. This can 
be done by examining SMDs, variance ratios, and other distributional statistics 
in the matched sample, or using plots such as empirical cumulative distribution 
functions. In the next subsection, we illustrate how propensity score matching 
can achieve covariate balance between plays where crossing was attempted relative 
to not crossing across the whole population of plays (estimand: ATE) and in the 
subpopulation defined by a cross being taken (estimand: ATT).

\subsection{Propensity score matching \label{sec:psm}}

Propensity score matching is a statistical technique widely used in 
observational studies to reduce or eliminate confounding and estimate causal 
effects by mimicking the conditions of a randomized controlled trial. The method 
involves pairing the plays where a cross was taken and where it was not with 
similar propensity scores. The success of propensity score matching depends on 
accurate specification of the propensity score model; misspecification can 
result in imbalanced groups and biased estimates. Various algorithms are 
available to match cross and no-cross plays, differing in how they measure 
similarity or distance. Common approaches include exact matching, nearest 
neighbor matching, and caliper matching \citep{austin2011introduction}. Matching 
can be performed on a one-to-one basis or extended to one-to-many matching 
configurations, and can be performed with or without replacement. Decisions 
concerning the closeness of the matching and whether to allow matching with 
replacement affect both the bias and variability of the resulting estimator. 
Requiring a very close match reduces covariate imbalance but may come at the 
cost of discarding some units that cannot be matched, inflating variance and 
possibly changing the target population. Similarly, matching with replacement 
may allow for closer matches, but could inflate variance due to the re-use of 
some units (plays) multiple times \citep{smith2005does}.

\subsubsection{Matching for estimation of the ATE}

When the causal effect of interest is the Average Treatment Effect, the target 
population includes all crossing-opportunities, whether they involved crossing 
or not. One-to-one matching for ATE matches cross plays to no-cross-plays 
\textit{and} no-cross-plays to cross-plays, creating a symmetric match structure 
to balance covariates across both groups. In effect, for each play $i$ where a 
cross was observed so that $Y_i=Y_i(1)$, we select a play under a similar 
situation - similar as defined by $\bX$ and summarized by the scalar summary 
of $\bX$ given by the propensity score - so as to use the outcome of the 
matched play, say $i'$ to `impute' $Y_i(0) = Y_{i'}$. Matching is typically 
performed with replacement, allowing a no-cross-play to be matched to multiple 
cross-plays, which can result in duplicate matches.

\subsubsection{Matching for estimation of the ATT}

When the causal effect of interest is the Average Treatment Effect on the 
Treated, the target population consists solely of the plays where a cross was 
taken, focusing on estimating the cross effect conditional on a cross actually 
being taken. One-to-one matching for ATT pairs each cross-play with the closest 
no-cross-play, leaving unmatched no-cross-plays unused in the analysis. The 
resulting distribution of $\bX$ in this subset of the population for which 
$Z=1$ differs from the marginal population of play situations. This distinction 
is consequential: when the propensity score distributions of cross and no-cross 
plays differ substantially, the ATE and ATT matching procedures marginalize over 
different regions of the covariate space, potentially yielding different causal 
estimates. We explore this further in Section~\ref{sec:simulation}.

\subsubsection{Implementation of matching in the crossing analysis 
\label{implementation}}

For our analysis, we used the \texttt{Matching} R package 
\citep{sekhon2011multivariate} to perform one-to-one nearest-neighbor 
propensity score matching. Nearest-neighbor matching pairs each cross-play with 
the no-cross-play whose propensity score is closest and - when the ATE is the 
estimand - also pairs each play without a cross to the closest (in propensity 
score) play where a cross was taken. We implemented this procedure with replacement to reduce bias, recognizing that it may increase variance \citep{austin2011introduction}. In our setting, where the number of cross-plays (5,614) is considerably smaller than the number of no-cross-plays (44,936), matching with replacement is particularly important as it allows each cross-play to 
be matched to its closest control regardless of whether that control 
has already been used, reducing bias at the cost of potentially 
inflating variance. Caliper matching is another commonly-used 
approach, however we chose not to use it in order to avoid discarding plays that 
lacked suitable matches. In addition, we allowed ties among matched pairs: if 
more than one no-cross-play had identical or nearly identical propensity scores 
to a given cross-play, all such no-cross-plays were retained. This strategy 
helps mitigate underestimation of the outcome variance that can occur when ties 
are not allowed \citep{sekhon2011multivariate}.

We evaluated covariate balance between plays with and without crossing using the SMD. We report both the Abadie--Imbens standard error and a bootstrapped 
standard error. The Abadie-Imbens standard error is derived assuming a known 
rather than estimated propensity score and hence may not accurately reflect the 
uncertainty in estimation, although it is widely used \citep{abadie2016matching}. 
Specifically, using the estimated propensity score as if it were true can 
produce conservative confidence intervals for the ATE, while its impact on the ATT 
remains unclear. We also include a bootstrapped standard error, given its common 
application in practice \citep{caliendo2008some}, despite evidence that 
bootstrap methods can fail under nearest-neighbor matching with replacement 
\citep{abadie2008failure}.

\section{Simulation study: when do ATE and ATT diverge?}
\label{sec:simulation}

\subsection{Motivation}

As discussed in Section~\ref{analysis_methods}, the ATE and ATT matching 
procedures marginalize over different regions of the covariate space - the 
full population of plays for the ATE, and the subpopulation of plays where a 
cross was taken for the ATT. When the propensity score distributions of cross 
and no-cross plays are similar, these two regions largely coincide, and the ATE 
and ATT will be close in value. When they differ substantially, however, the 
two estimands can diverge meaningfully. To illustrate this mechanism, we 
construct two simulation scenarios using covariates drawn from the empirical 
dataset, allowing us to work with realistic covariate distributions while 
controlling the data-generating process precisely. Crucially, because the 
potential outcomes are generated by construction, the true values of the ATE 
and ATT are known, enabling a direct evaluation of whether propensity score 
matching successfully recovers each estimand.

The two scenarios are designed to represent contrasting situations: Scenario A 
reproduces the high-overlap structure of the real data, where the propensity 
score distributions of cross and no-cross plays are similar and the ATE and ATT 
are expected to coincide; Scenario B is engineered to exhibit severe confounding 
and low overlap in propensity scores, forcing the ATE and ATT to diverge. In 
both scenarios, we apply the same 1:1 nearest-neighbor propensity score matching 
procedure described in Section~\ref{sec:psm} and evaluate whether matching 
successfully recovers the true estimands. Pre- and post-matching covariate 
balance tables for both scenarios are provided in the Appendix 
(Tables~\ref{tab:balance_scenA} and~\ref{tab:balance_scenB}).

\subsection{Scenario A: similar propensity score distributions 
(\texorpdfstring{ATE $\approx$ ATT}{ATE approximately equal to ATT})}

\subsubsection{Data generating mechanism}

Let $T \in \{0, 1\}$ denote the binary variable representing the crossing 
attempt and $Y \in \{0, 1\}$ denote the binary outcome representing the shot 
attempt. Let $Y^{(0)}$ and $Y^{(1)}$ represent the potential outcomes under no 
cross and cross, respectively. We sample $N = 10{,}000$ observations with 
replacement from the empirical dataset, extracting two continuous covariates 
which are standardized (mean zero, unit variance) prior to simulation:

\begin{itemize}
    \item $X_1$: Standardized distance of the ball carrier to the opponent's 
    goal line.
    \item $X_2$: Standardized offensive-defensive player ratio in the box.
\end{itemize}

To induce moderate confounding, the crossing action $T$ is generated as a 
Bernoulli random variable with probabilities derived from the following logistic 
regression model:

\begin{equation}
    \text{logit}(P(T = 1 \mid X_1, X_2)) = -2.1 + 0.8X_1 + 0.6X_2.
    \label{eq:ps_scenA}
\end{equation}

The coefficients are strictly non-zero, ensuring that crossing is dependent on 
the spatial covariates, thereby necessitating covariate adjustment. However, 
their moderate magnitudes prevent strict separation, ensuring sufficient overlap 
in the propensity score distributions between cross and no-cross plays, as 
illustrated in Figure~\ref{fig:true_ps_overlap} in the Appendix. The observed outcome $Y$ is 
generated via Bernoulli trials, conditional on the covariates and the crossing 
status $T$:
\begin{equation}
    \text{logit}(P(Y = 1 \mid T, X_1, X_2)) = -3.9 + 0.05X_1 + 0.03X_2 + 1.3T.
    \label{eq:outcome_scenA}
\end{equation}

Under the potential outcomes framework, the unobservable potential outcomes 
$Y^{(0)}$ and $Y^{(1)}$ are recovered by evaluating this structural equation at 
$T = 0$ and $T = 1$, respectively. This specification inherently satisfies the 
causal consistency assumption, where $Y = TY^{(1)} + (1-T)Y^{(0)}$. Because the 
treatment effect is homogeneous, the coefficient on $T$ does not interact with 
any covariate, so the true ATE and ATT converge to identical theoretical values 
by construction.

\subsubsection{Results}

\begin{figure}
    \small\sf\centering
    \includegraphics[width=0.9\linewidth]{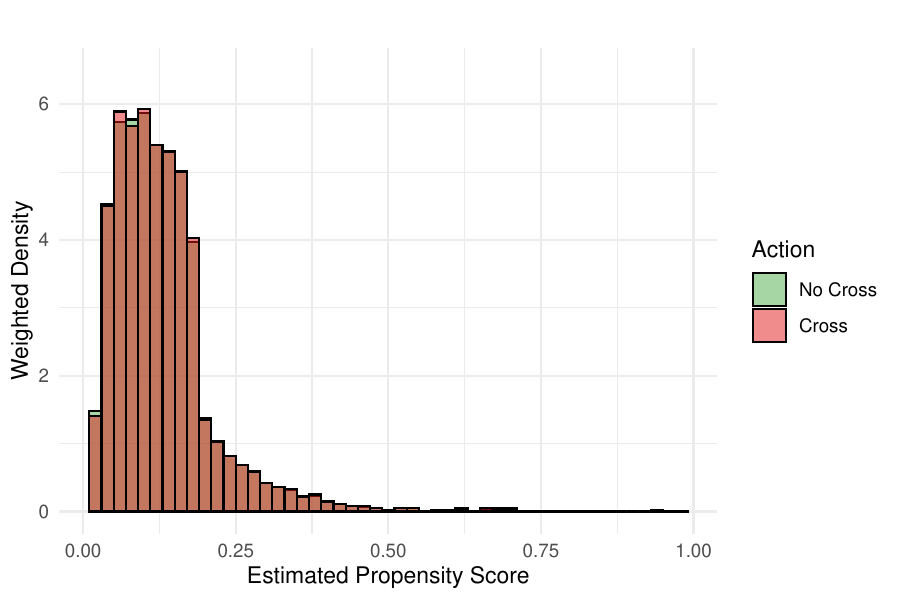}
    \includegraphics[width=0.9\linewidth]{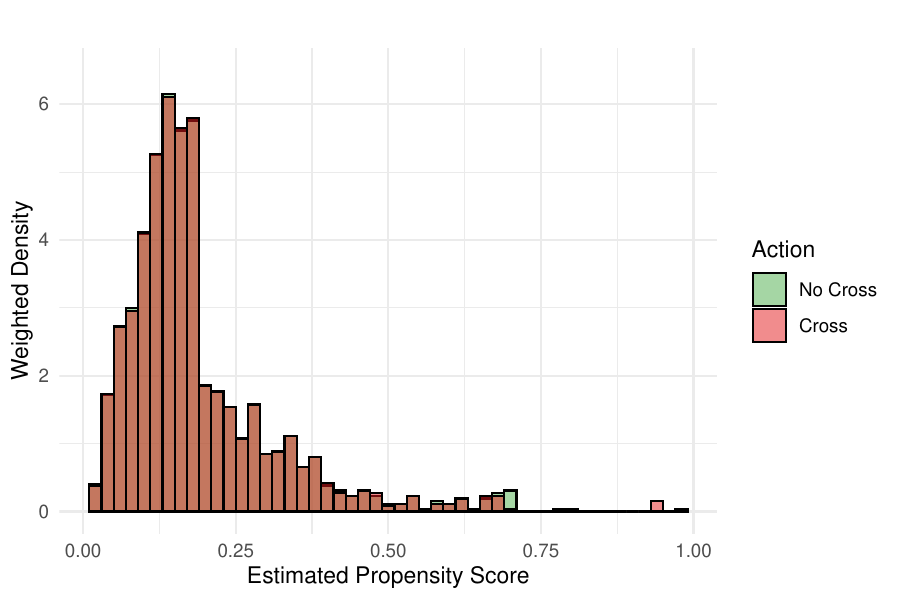}
    \caption{Weighted post-matching propensity score distributions for (a) ATE (top) and (b) ATT (bottom) in Scenario A.}
    \label{fig:ps_overlap_post_matching_weighted_A}
\end{figure}

Figure~\ref{fig:true_ps_overlap} in the Appendix shows the distributions of the true propensity 
scores prior to any estimation, confirming that the cross and no-cross plays 
occupy largely the same region of the propensity score space. Following 1:1 
nearest-neighbor matching with replacement, 
Figure~\ref{fig:ps_overlap_post_matching_weighted_A} displays the weighted 
post-matching propensity score distributions for the ATE (top) and ATT (bottom) 
analytic samples. Since matching is performed with replacement, some untreated 
plays are matched to multiple treated plays; the frequency weights reflect this 
reuse, accurately representing the covariate distribution that each estimator 
marginalizes over. The near-identical distributions across both panels confirm 
that matching for the ATE and ATT marginalizes over the same region of the 
covariate space in Scenario A, which is a direct consequence of the high overlap 
in propensity scores.

Table~\ref{tab:estimand_results_both} (Scenario A panel) reports the estimated 
causal effects against the true parameters. Because Scenario A features high propensity score overlap 
and a homogeneous treatment effect, the true ATE and ATT are nearly identical 
(0.0503 and 0.0515, respectively), and the matching estimator closely recovers 
both parameters (estimated ATE: 0.0520, Abadie--Imbens SE: 0.0082, bootstrap 
SE: 0.0093; estimated ATT: 0.0539, Abadie--Imbens SE: 0.0081, bootstrap SE: 
0.0078). This scenario illustrates that when the propensity score distributions 
overlap substantially, both estimands are estimable with comparable precision 
and yield similar substantive conclusions - the choice of estimand is 
consequential in principle but immaterial in practice.

\begin{table}
    \small\sf\centering
\caption{Comparison of true vs.\ estimated causal estimands (Risk Difference) 
for Scenarios A and B. SE (A-I): Abadie--Imbens standard error.}
    \label{tab:estimand_results_both}
    \begin{tabular}{llcccc}
        \toprule
        & \textbf{Estimand} & \textbf{True Parameter} & \textbf{Estimated Effect} 
        & \textbf{SE (A-I)} & \textbf{SE (Bootstrap)} \\
        \midrule
        \multicolumn{6}{l}{\textit{Scenario A}} \\
        \midrule
        & ATE & 0.0503 & 0.0520 & 0.0082 & 0.0093 \\
        & ATT & 0.0515 & 0.0539 & 
        0.0081 & 0.0078 \\
        \midrule
        \multicolumn{6}{l}{\textit{Scenario B}} \\
        \midrule
        & ATE & 0.0684 & 0.0740 & 0.0148 & 0.0085 \\
        & ATT & 0.1045 & 0.0962 & 
        0.0176 & 0.0081 \\
        \bottomrule
    \end{tabular}
\end{table}

\subsection{Scenario B: separated propensity score distributions 
(\texorpdfstring{ATE $\neq$ ATT}{ATE not equal to ATT})}

\subsubsection{Data generating mechanism}

As in Scenario A, we draw $N = 10{,}000$ observations with replacement from the 
empirical dataset, extracting the same standardized spatial covariates $X_1$ 
(distance to endline) and $X_2$ (offensive-defensive ratio). To induce severe 
confounding and low overlap in the propensity scores, we inflate the magnitude 
of the coefficients in the propensity score model:

\begin{equation}
    \text{logit}(P(T = 1 \mid X_1, X_2)) = -1.5 + 2.5X_1 + 2X_2.
    \label{eq:ps_scenB}
\end{equation}

Consequently, the distributions of the true propensity scores for those who 
crossed and those who did not strictly separate, generating a near violation of 
the positivity assumption, as illustrated in Figure~\ref{fig:true_ps_overlap_B} in  the Appendix. 
To further force the ATE and ATT to diverge, we introduce a heterogeneous 
treatment effect by including an interaction term between the crossing action 
and the distance to the goal line:
\begin{align}
    \text{logit}(P(Y = 1 \mid T, X_1, X_2)) &= -3.9 + 0.2X_1 - 0.1X_2 +\nonumber\\
    &  1.3T 
    + 0.6(T \cdot X_1).
    \label{eq:outcome_scenB}
\end{align}
Under this parameterization, the main effect ($\beta_T = 1.3$) anchors the 
population-level ATE, while the interaction term ($\beta_{T \cdot X_1} = 0.6$) 
interacts with the highly selected spatial distribution of those who crossed to 
drive the ATT significantly higher. That is, crossing tends to occur when the 
ball carrier is closer to the goal line - precisely the situations where the 
interaction term amplifies the treatment effect - so the effect among those 
who actually crossed is larger than the effect averaged across the full 
population.

\subsubsection{Results}

\begin{figure}
    \small\sf\centering
    \includegraphics[width=0.9\linewidth]{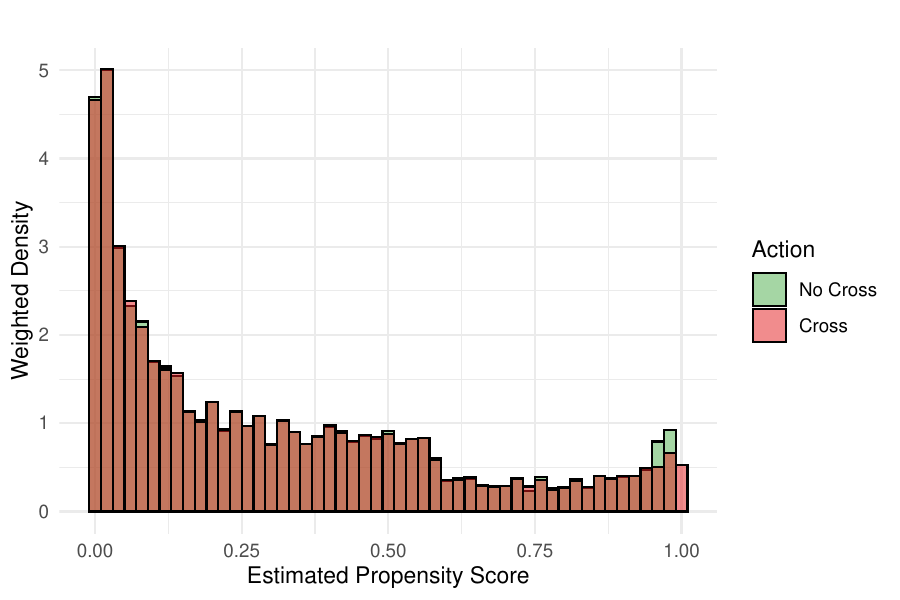}
    \includegraphics[width=0.9\linewidth]{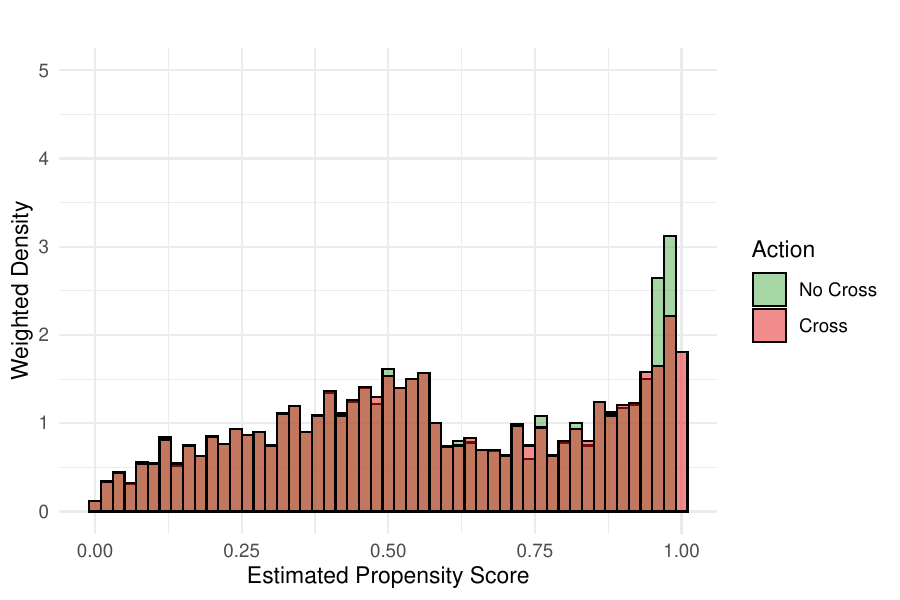}
    \caption{Weighted post-matching propensity score distributions for (a) ATE (top) and (b) ATT (bottom) in Scenario B.}
    \label{fig:ps_overlap_post_matching_weighted_B}
\end{figure}

In the Appendix, Figure~\ref{fig:true_ps_overlap_B} shows the distributions of the true 
propensity scores prior to any estimation. In contrast to Scenario A, the cross 
and no-cross plays now occupy largely distinct regions of the propensity score 
space, confirming the near violation of the positivity assumption induced by the 
data generating mechanism. Following 1:1 nearest-neighbor matching with 
replacement, Figure~\ref{fig:ps_overlap_post_matching_weighted_B} displays the 
weighted post-matching propensity score distributions for the ATE (top) and 
ATT (bottom) analytic samples. Unlike Scenario A, the two panels now exhibit 
markedly different distributions: the ATE matching marginalizes over a broader, 
more uniform region of the covariate space, while the ATT matching concentrates 
on the higher propensity score region characteristic of plays where a cross was 
actually taken. This divergence in the effective covariate distributions, 
combined with the heterogeneous treatment effect specified in the outcome model, 
is what drives the ATE and ATT apart.

Table~\ref{tab:estimand_results_both} (Scenario B panel) compares the true parameters against the 
empirical estimates. As dictated by the data generating mechanism, the true ATE 
and ATT diverge substantially (0.0684 and 0.1045, respectively). The matching 
estimator closely approximates these distinct targets (estimated ATE: 0.0740, 
Abadie--Imbens SE: 0.0148, bootstrap SE: 0.0085; estimated ATT: 0.0962, 
Abadie--Imbens SE: 0.0176, bootstrap SE: 0.0081). This scenario illustrates 
that when crosses tend to occur in situations that are particularly favorable 
for shot creation - that is, when treatment selection correlates with effect 
heterogeneity - the ATT captures a meaningfully larger effect than the ATE. 
In such settings, the choice of estimand is not merely a technical detail but 
a substantive decision that shapes the conclusions drawn from the analysis.

\subsection{Key takeaway}

Taken together, the two scenarios demonstrate that the divergence between ATE 
and ATT is driven by two interacting factors: the degree of separation in 
propensity score distributions between treated and untreated units, and the 
presence of treatment effect heterogeneity. When propensity scores overlap 
substantially (Scenario A), matching for the ATE and ATT marginalizes over 
nearly the same covariate distribution, and the two estimands converge. When 
propensity scores are separated (Scenario B), matching for the ATE and ATT 
marginalizes over different covariate distributions, and in the presence of 
effect heterogeneity, the estimands diverge. In both scenarios, propensity 
score matching closely recovers the true estimands, providing empirical support 
for the methodology in these settings. In Section~\ref{sec:casestudy}, we 
return to the soccer case study and show that the real data resembles Scenario 
A - a finding that, in light of the simulation, is consistent with the high 
overlap in propensity scores between cross and no-cross plays, though we cannot 
rule out other contributing factors.

\section{Case study: crossing in soccer}
\label{sec:casestudy}

\subsection{Data}

The dataset includes detailed tracking data on crossing opportunities and player 
movements in potential crossing zones from nearly all 240 matches of the 2019 
Chinese Super League season, spanning all teams in the league. As described in Section~\ref{sec:Intro}, the dataset captures 
50,550 crossing opportunities across nearly all 240 matches of 
the 2019 Chinese Super League season. We take as our outcome the binary indicator of whether a shot 
resulted from the play. Several potential confounders derived from 
spatiotemporal tracking data are included, as described in 
Section~\ref{sec:framework}.

Table~\ref{tab:desc} presents descriptive statistics stratified by crossing 
status. There is considerable covariate imbalance between the cross and no-cross 
groups, with cross attempts tending to occur when the ball carrier was closer to 
the opponent's goal line (mean 15.32 vs.\ 21.87 meters), when the 
offensive-defensive player ratio in the box was more favorable (mean 0.43 
vs.\ 0.22), and when a teammate was closer to the opponent's goal line (mean 
19.14 vs.\ 24.32 meters). Defenders were also more likely to attempt crosses 
(37.4\%) compared to forwards (17.1\%). All covariates except defender closing 
time and the ten minute warning indicator showed SMDs exceeding 10\%, 
underscoring the need for confounding adjustment.

\begin{table}
    \small\sf\centering
    \caption{Descriptive statistics and standardized absolute mean differences 
    of covariates stratified by cross attempt status in the 2019 CSL dataset.}
    \label{tab:desc}
    \resizebox{\textwidth}{!}{%
    \begin{tabular}{lcccc}
       \toprule
        & \textbf{Overall} & \textbf{No Cross (0)} & \textbf{Cross (1)} & 
        \textbf{SMD (\%)} \\
        \midrule
        Sample size ($n$) & 50,550 & 44,936 & 5,614 & \\
        Score differential (mean (SD)) & -0.15 (1.22) & -0.13 (1.22) & 
        -0.29 (1.13) & 12.96 \\
        Defender closing time (mean (SD)) & 1.43 (2.59) & 1.44 (2.61) & 
        1.41 (2.46) & 0.99 \\
        Distance to sideline (mean (SD)) & 6.41 (3.81) & 6.31 (3.82) & 
        7.20 (3.61) & 24.04 \\
        Teammate distance to goal line (mean (SD)) & 23.75 (9.30) & 
        24.32 (9.23) & 19.14 (8.56) & 58.23 \\
        Distance to endline (mean (SD)) & 21.15 (9.17) & 21.87 (8.96) & 
        15.32 (8.75) & 74.03 \\
        Offensive-defensive ratio in box (mean (SD)) & 0.25 (0.26) & 
        0.22 (0.25) & 0.43 (0.21) & 87.23 \\
        Position of crosser (\%) & & & & 38.40 \\
        \quad Forward & 14,247 (28.2) & 13,287 (29.6) & 960 (17.1) & \\
        \quad Midfielder & 24,082 (47.6) & 21,530 (47.9) & 2,552 (45.5) & \\
        \quad Defender & 12,221 (24.2) & 10,119 (22.5) & 2,102 (37.4) & \\
        Ten minute warning (=1) (\%) & 14,237 (28.2) & 12,684 (28.2) & 
        1,553 (27.7) & 1.26 \\
        \bottomrule
    \end{tabular}}
\end{table}

\subsection{Results}

\begin{figure}
    \small\sf\centering
    \includegraphics[width=1\textwidth]{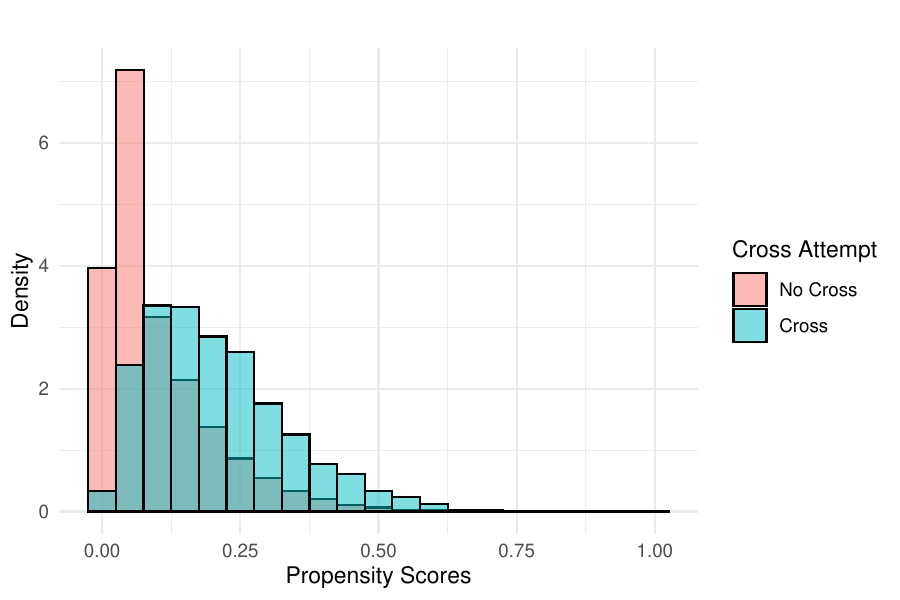}
    \caption{The distributions of the propensity scores for the action of cross 
    (cyan) or no-cross (salmon) prior to matching.}
    \label{fig:pre_match}
\end{figure}

We used logistic regression to fit the propensity score model; the output is 
provided in Table~\ref{tab:ps_model} in the Appendix. The substantial overlap in propensity scores visible in 
Figure~\ref{fig:pre_match} provides empirical support for the 
positivity assumption in this dataset, suggesting that for most 
covariate combinations, both crossing and not crossing were 
plausible actions during the 2019 CSL season. Despite the substantial pre-matching imbalance, the 
propensity score distributions of cross and no-cross plays exhibit considerable 
overlap across the full range of scores, a feature we return to in 
Section~\ref{sec:interp}.

\begin{figure}
    \small\sf\centering
    \includegraphics[width=1\linewidth]{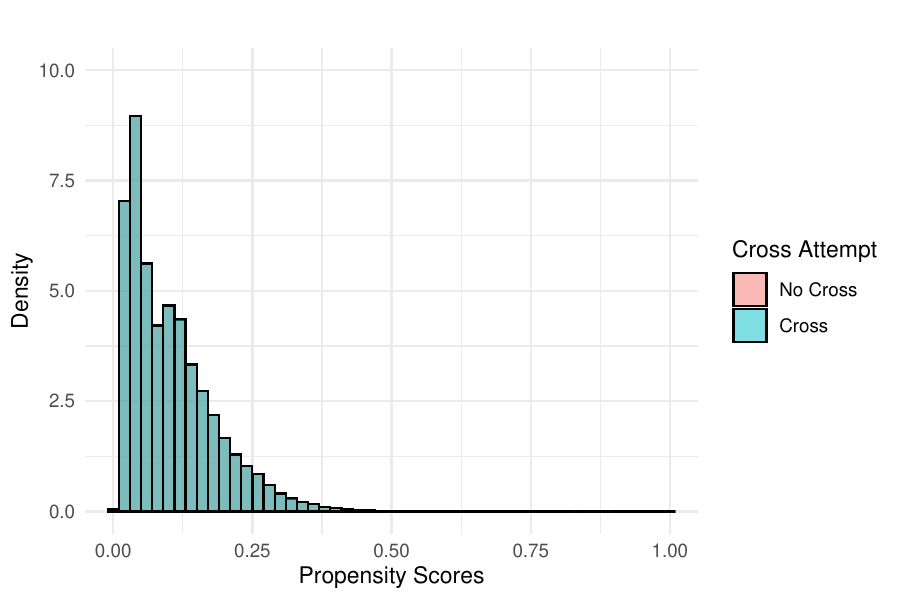}
    \includegraphics[width=1\linewidth]{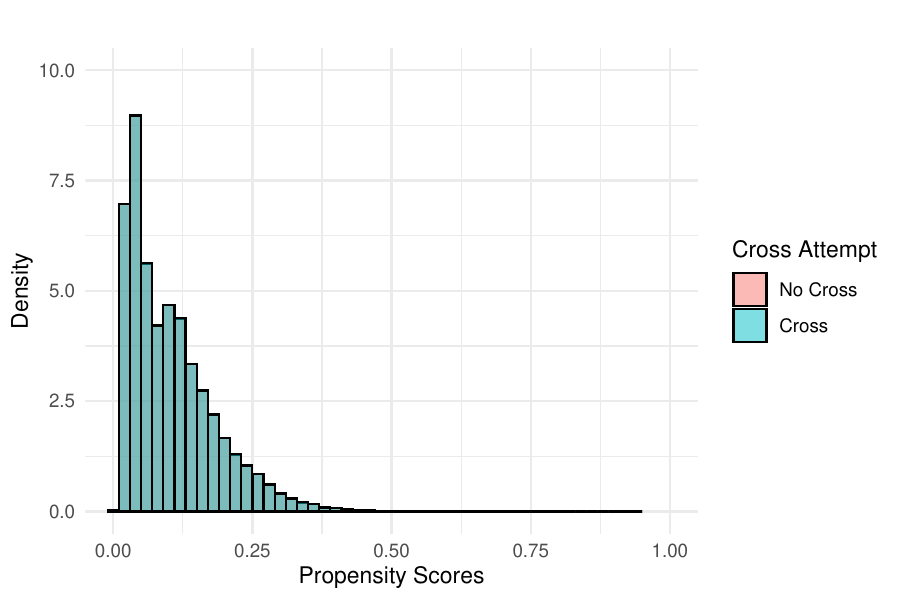}
    \caption{Post-matching propensity score distributions when (a) ATE (top) is the estimand and (b) ATT (bottom) is the estimand.}
    \label{fig:post_match}
\end{figure}

Following 1:1 nearest-neighbor matching with replacement, covariate balance was 
successfully achieved for both the ATE and ATT analytic samples. 
Tables~\ref{tab:balance_ATE} and~\ref{tab:balance_ATT} show the post-matching 
covariate distributions for the ATE and ATT samples respectively, with all SMDs 
reduced well below the 10\% threshold. The distributions of the propensity 
scores after matching are shown in Figure~\ref{fig:post_match} for the ATE 
(top) and ATT (bottom), with the two action groups exhibiting near-identical 
distributions in both cases.

\begin{table}
    \small\sf\centering
    \caption{Post-matching covariate balance stratified by cross status when 
    the estimand is ATE. All SMDs are below the 10\% threshold, indicating 
    successful covariate balance following matching.}
    \label{tab:balance_ATE}
    \begin{tabular}{lccc}
        \toprule
        & \textbf{No Cross (0)} & \textbf{Cross (1)} & \textbf{SMD (\%)} \\
        \midrule
        Sample size ($n$) & 793,158 & 793,158 & \\
        Score differential (mean (SD)) & -0.17 (1.21) & -0.18 (1.12) & 1.06 \\
        Defender closing time (mean (SD)) & 1.45 (2.59) & 1.64 (2.92) & 6.63 \\
        Distance to sideline (mean (SD)) & 6.48 (3.82) & 6.36 (3.42) & 3.34 \\
        Teammate distance to goal line (mean (SD)) & 23.04 (8.76) & 
        23.44 (9.25) & 4.52 \\
        Distance to endline (mean (SD)) & 20.58 (8.62) & 21.09 (8.50) & 5.88 \\
        Offensive-defensive ratio in box (mean (SD)) & 0.26 (0.24) & 
        0.28 (0.23) & 8.92 \\
        Position of crosser (\%) & & & 0.68 \\
        \quad Forward & 209,063 (26.4) & 211,373 (26.6) & \\
        \quad Midfielder & 393,151 (49.6) & 391,078 (49.3) & \\
        \quad Defender & 190,944 (24.1) & 190,707 (24.0) & \\
        Ten minute warning (=1) (\%) & 221,639 (27.9) & 228,175 (28.8) & 1.83 \\
        \bottomrule
    \end{tabular}
\end{table}

\begin{table}
    \small\sf\centering
    \caption{Post-matching covariate balance stratified by cross status when 
    the estimand is ATT. All SMDs are below the 10\% threshold, indicating 
    successful covariate balance following matching.}
    \label{tab:balance_ATT}
    \begin{tabular}{lccc}
        \toprule
        & \textbf{No Cross (0)} & \textbf{Cross (1)} & \textbf{SMD (\%)} \\
        \midrule
        Sample size ($n$) & 394,032 & 394,032 & \\
        Score differential (mean (SD)) & -0.17 (1.21) & -0.19 (1.12) & 1.07 \\
        Defender closing time (mean (SD)) & 1.45 (2.59) & 1.64 (2.93) & 6.63 \\
        Distance to sideline (mean (SD)) & 6.48 (3.82) & 6.36 (3.42) & 3.28 \\
        Teammate distance to goal line (mean (SD)) & 23.01 (8.75) & 
        23.42 (9.24) & 4.53 \\
        Distance to endline (mean (SD)) & 20.56 (8.62) & 21.07 (8.50) & 5.96 \\
        Offensive-defensive ratio in box (mean (SD)) & 0.26 (0.24) & 
        0.28 (0.23) & 8.95 \\
        Position of crosser (\%) & & & 0.67 \\
        \quad Forward & 103,573 (26.3) & 104,682 (26.6) & \\
        \quad Midfielder & 195,464 (49.6) & 194,419 (49.3) & \\
        \quad Defender & 94,995 (24.1) & 94,931 (24.1) & \\
        Ten minute warning (=1) (\%) & 110,081 (27.9) & 113,390 (28.8) & 1.86 \\
        \bottomrule
    \end{tabular}
\end{table}

Estimation of the ATE and ATT proceeds by computing a simple difference of mean 
outcomes in each action group following matching. The results are summarized in 
Table~\ref{tab:estimates}. The ATE was estimated to be 0.033, implying that 
crossing the ball increases the probability of a resulting shot by 3.3\% on 
average across all plays. The ATT was estimated to be 0.035, implying that 
among plays where a cross was actually attempted, crossing increased the 
probability of a shot by 3.5\% on average. The Abadie--Imbens standard errors 
are 0.005 and 0.003 for the ATE and ATT respectively, and the bootstrapped 
standard errors, based on 1000 resamples, are 0.0052 and 
0.0033. These estimates should be interpreted in the context of the 2019 Chinese Super League season and may not generalize to other leagues 
or styles of play.

\begin{table}
    \small\sf\centering
     \caption{Estimated causal effects of crossing on shot probability in the 2019 CSL dataset.}
    \label{tab:estimates}
    \begin{tabular}{lccc}
        \toprule
        \textbf{Estimand} & \textbf{Estimate} & \textbf{SE (A-I)} & 
        \textbf{SE (Bootstrap)} \\
        \midrule
        ATE & 0.033 & 0.005 & 0.0052 \\
        ATT & 0.035 & 0.003 & 
        0.0033 \\
        \bottomrule
    \end{tabular}
\end{table}

\subsection{Interpreting the findings in light of the simulation}
\label{sec:interp}

The ATE and ATT estimates are nearly identical (0.033 and 0.035 respectively), 
a finding that warrants explanation in light of the simulation study in 
Section~\ref{sec:simulation}. As demonstrated in Scenario A, when the 
propensity score distributions of cross and no-cross plays overlap 
substantially, the ATE and ATT matching procedures marginalize over essentially 
the same covariate distribution, causing the two estimands to converge. 
Figure~\ref{fig:pre_match} confirms that this is precisely the structure of the 
real data - the propensity scores of cross and no-cross plays occupy largely 
the same region of the score space, and Figure~\ref{fig:post_match} shows that 
the post-matching distributions are nearly identical across both estimands. The 
near-identical ATE and ATT estimates are therefore consistent with what the 
simulation predicted for this type of high-overlap setting, and should be 
interpreted as an informative finding in its own right rather than a failure to 
distinguish the two estimands. In this particular dataset, the effect of 
crossing on shot creation is relatively uniform across the population of 
crossing opportunities, and the choice of estimand is immaterial in practice.

\section{Discussion \label{sec:discussion}}

This study provides a tutorial on the importance of estimand choice in causal 
inference, illustrated through an analysis of crossing in soccer and a 
simulation study designed to elucidate the conditions under which the ATE and 
ATT diverge. Using data from nearly all 240 matches of the 2019 Chinese Super 
League season, propensity score matching yielded nearly identical ATE and ATT 
estimates (Section~\ref{sec:casestudy}), suggesting that crossing 
yields a modest but consistent increase in shot probability, 
regardless of the population over which the effect is averaged. The simulation 
study showed that this convergence is not coincidental: when propensity score 
distributions overlap substantially between treated and untreated units, as they 
do in this dataset, the ATE and ATT matching procedures marginalize over 
essentially the same covariate distribution, causing the two estimands to 
converge. When overlap is low and treatment effects are heterogeneous, as in 
Scenario B, the estimands diverge meaningfully, with substantively different implications for practice. 

From a practical standpoint, the near-identical ATE and ATT estimates suggest 
that the effect of crossing on shot creation is relatively uniform across the 
population of crossing opportunities in this dataset, and that players' 
decisions to cross do not appear to be concentrated in situations where crossing 
is particularly more or less effective than average. Coaches operating in a similar context to the 2019 Chinese Super 
League can leverage this information to inform training programs 
and tactical decision-making, though caution is warranted when 
extrapolating these findings to other leagues or levels of play. The extent to which these findings generalize may depend on a team's 
tactical approach, for instance, teams that rely heavily on wide 
play may see different effects of crossing than those that prefer 
central build-up, a distinction explored in the team playing styles 
literature \citep{vecer2014crossing, sarkar2018paradox}. More 
broadly, causal assessments of in-game tactics may in general be better answered 
by the ATT, whereas questions about the effect of a training regime or other 
intervention that could reasonably be deployed on a wider scale are often more 
naturally answered via the ATE. When the two estimands diverge, this divergence 
is itself informative  suggesting that the tactic is being selectively 
deployed in situations where it is particularly effective, and pointing toward a 
more in-depth investigation of the conditions under which the tactic is most 
beneficial. Analytic strategies for identifying such subgroups, such as those 
from the framework of precision interventions or individualized treatment rules, 
may also be relevant for future work in this area 
\citep{wallace2014personalizing}. The causal inference framework illustrated here can be directly applicable 
to analogous tactical decisions in other sports, for instance, the 
effectiveness of serve strategy in tennis or penalty corner execution 
in field hockey, wherever observational tracking data are available 
and the estimand of interest can be clearly defined.

The choice of estimand also has practical implications for the analytic tools 
available. Several analytic strategies exist that avoid matching altogether, 
relying instead on weighting or highly flexible modeling 
\citep{goetghebeur2020formulating}, and software exists to implement doubly 
robust alternatives \citep{zetterqvist2015doubly, zhong2021aipw}. However, many 
of these approaches are primarily implemented for the ATE rather than the ATT, 
which further underscores the importance of clearly specifying the estimand of 
interest before selecting an analytic approach, as the choice of estimand can 
constrain the methods available for estimation. It is also worth noting that in 
\citet{dona2023causal, epasinghege2024causal} and \citet{wu2021contextual}, the 
authors use the term ATE as the causal estimand whereas their measure would 
conventionally be described as the ATT. Regression adjustment within a 
propensity-score matched sample is another approach to address residual 
confounding and reduce bias due to covariate imbalance 
\citep{rubin2000combining}, and can enhance the precision of causal effect 
estimates by making them doubly robust \citep{austin2017double}. Evidence 
suggests that covariate adjustment is most beneficial for variables with 
standardized absolute mean differences exceeding 10\%, and prioritizing 
adjustment for these covariates is recommended to mitigate imbalance 
\citep{nguyen2017double}.

There are several limitations to this analysis. First, the no interference 
assumption may not hold in soccer, as teams may adjust their strategies in 
response to earlier plays and tactical interdependencies between players are 
common. The use of multiple confounders may help reduce the impact of any 
previous plays on a given play, but this remains a limitation. Second, although 
propensity score matching effectively improved covariate balance, SMDs that 
remained near 10\% after matching may result in residual confounding. Propensity 
score matching relies heavily on having measured all relevant confounders and on 
correctly specifying the propensity score model, and unmeasured confounders such as the distance between the last defender and the byline, the opponent's defensive strategy, or the shooting proficiency of attacking players may bias the estimates. Third, although the dataset spans nearly all 240 
matches of the 2019 Chinese Super League season, the results may not generalize 
to other leagues or styles of play. Propensity score distributions in other leagues may differ from those observed here, potentially leading to greater divergence between the ATE and ATT. Indeed, \citet{vecer2014crossing} found that crossing has a negative impact on scoring in top European leagues such as the Premier League 
and Bundesliga, suggesting that the effect of crossing may vary 
substantially across leagues of different technical levels. Extending this analysis to data from multiple leagues would therefore provide a more comprehensive understanding of the causal impact of crossing in soccer.

Future work could explore analyzing the data within the framework of dynamic 
treatment regimes, which offers the potential to identify specific strategies 
valuable for planning purposes. Incorporating decision rules could lead to 
strategies that are not only more interpretable but also more parsimonious, 
enhancing their practical applicability for in-game play. More broadly, the causal inference framework illustrated here, including the careful choice between the ATE and ATT, is directly applicable to analogous tactical decisions in other sports, such as serve strategy in tennis or penalty corner execution in field hockey, wherever observational tracking data are available. The insights found in 
this work underscore the value of integrating causal inference techniques into 
sports analytics to better understand and improve team strategies.

\section{Conclusion}

This paper provided a step-by-step tutorial on framing causal questions 
precisely and applying appropriate statistical methods to answer them, 
illustrated through a case study of crossing in soccer. By demonstrating both empirically and through simulation the conditions under which the choice between the ATE and ATT matters, we hope to equip practitioners in sports analytics with the tools to make rigorous and informed methodological choices tailored to their specific research goals. Understanding and clearly framing 
analytic questions through a causal lens is essential for ensuring the 
reliability and interpretability of conclusions drawn from observational sports 
data.



\bibliography{Soccer_Causal}
\bibliographystyle{SageH}

\input{Appendix}

\end{document}

%% file: Appendix.tex
\newpage
\appendix
\renewcommand{\thesection}{\Alph{section}}
\renewcommand{\thesubsection}{\Alph{section}.\arabic{subsection}}
\renewcommand{\thetable}{\thesection.\arabic{table}}
\renewcommand{\thefigure}{\thesection.\arabic{figure}}
\setcounter{table}{0}
\setcounter{figure}{0}

\begin{center}
    {\LARGE\textbf{Appendix}}\\
    \vspace{1em}
    \end{center}
\vspace{1em}

\section{Additional simulation results}
\subsection{True propensity score distribution for Scenario A}

\begin{figure}[h!]
    \small\sf\centering
   \includegraphics[width=0.9\textwidth]{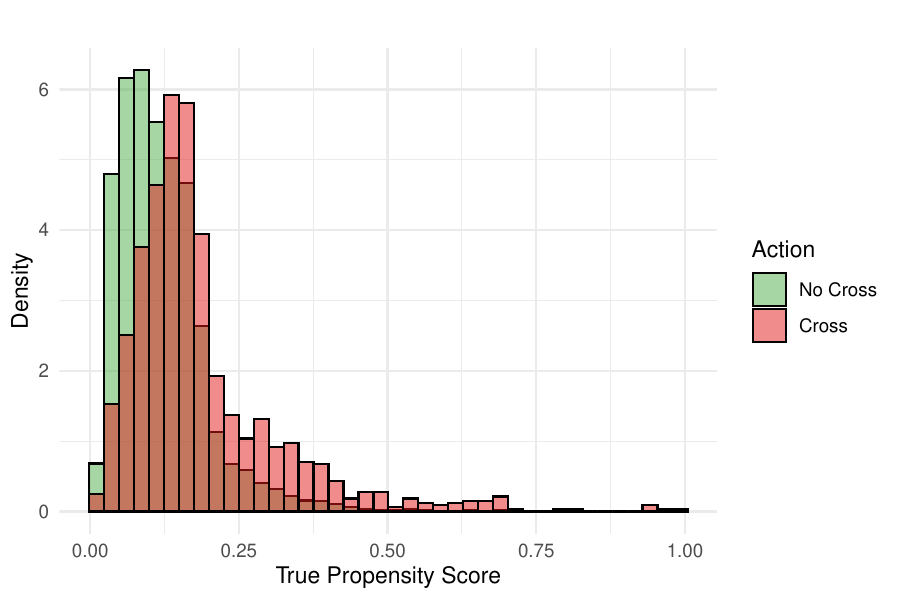}
    \caption{True propensity score distributions derived from the data generating mechanism in Scenario A.}
    \label{fig:true_ps_overlap}
\end{figure}

\subsection{Covariate balance for Scenario A}
\begin{table}[h!]
    \small\sf\centering
\caption{Covariate balance before and after matching for Scenario A. 
All post-matching SMDs are below the 10\% threshold, indicating 
successful covariate balance following matching.}
    \label{tab:balance_scenA}
    \begin{tabular}{llccc}
        \toprule
        & \textbf{Covariate} & \textbf{No Cross ($T=0$)} & 
        \textbf{Cross ($T=1$)} & \textbf{SMD (\%)} \\
        \midrule
        \multicolumn{5}{l}{\textit{Pre-matching ($N_{\text{No Cross}} = 8{,}698$; $N_{\text{Cross}} = 1{,}302$)}} \\
        \midrule
        & Distance to Endline ($X_1$) & 20.66 (9.19) & 24.55 (7.87) & 45.44 \\
        & Offensive-Defensive Ratio ($X_2$) & 0.24 (0.24) & 0.32 (0.32) & 28.63 \\
        \midrule
        \multicolumn{5}{l}{\textit{Post-matching: ATE ($N = 56{,}709$ per group)}} \\
        \midrule
        & Distance to Endline ($X_1$) & 24.16 (8.59) & 24.23 (8.53) & 0.76 \\
        & Offensive-Defensive Ratio ($X_2$) & 0.18 (0.23) & 0.17 (0.23) & 1.00 \\
        \midrule
        \multicolumn{5}{l}{\textit{Post-matching: ATT ($N = 27{,}149$ per group)}} \\
        \midrule
        & Distance to Endline ($X_1$) & 24.59 (8.38) & 24.51 (8.35) & 1.00 \\
        & Offensive-Defensive Ratio ($X_2$) & 0.17 (0.23) & 0.17 (0.23) & 1.43 \\
        \bottomrule
    \end{tabular}
\end{table}

\FloatBarrier
\subsection{True propensity score distribution for Scenario B}
\begin{figure}[h!]
    \small\sf\centering
    \includegraphics[width=0.9\textwidth]{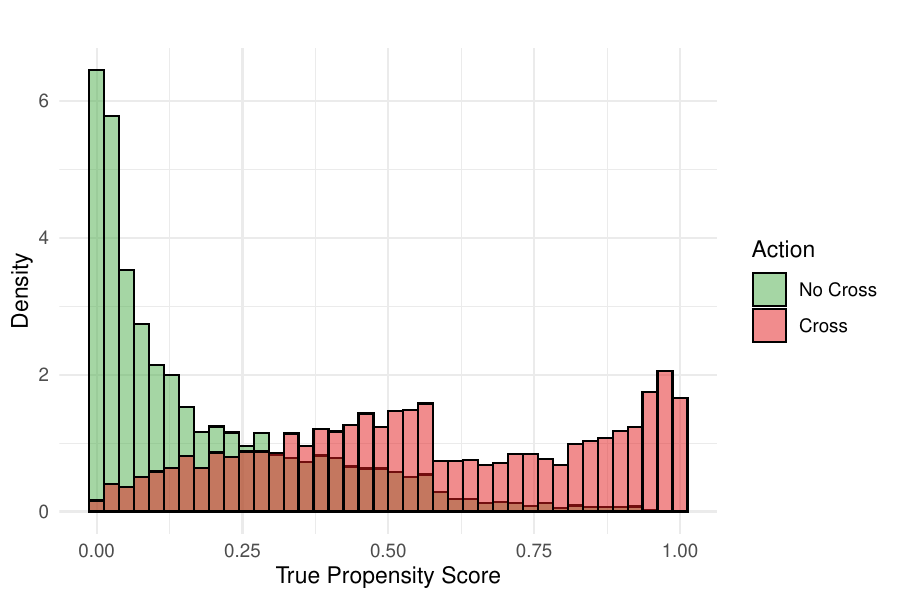}
    \caption{True propensity score distributions derived from the data generating mechanism in Scenario B.}
    \label{fig:true_ps_overlap_B}
\end{figure}

\clearpage
\subsection{Covariate balance for Scenario B}
\begin{table}[h!]
    \small\sf\centering
\caption{Covariate balance before and after matching for Scenario B. 
All post-matching SMDs are below the 10\% threshold, indicating 
successful covariate balance following matching.}
    \label{tab:balance_scenB}
    \begin{tabular}{llccc}
        \toprule
        & \textbf{Covariate} & \textbf{No Cross ($T=0$)} & 
        \textbf{Cross ($T=1$)} & \textbf{SMD (\%)} \\
        \midrule
        \multicolumn{5}{l}{\textit{Pre-matching ($N_{\text{No Cross}} = 7{,}061$; $N_{\text{Cross}} = 2{,}939$)}} \\
        \midrule
        & Distance to Endline ($X_1$) & 19.10 (9.25) & 26.12 (6.57) & 87.47 \\
        & Offensive-Defensive Ratio ($X_2$) & 0.21 (0.22) & 0.34 (0.30) & 51.67 \\
        \midrule
        \multicolumn{5}{l}{\textit{Post-matching: ATE ($N = 91{,}371$ per group)}} \\
        \midrule
        & Distance to Endline ($X_1$) & 25.35 (8.03) & 25.70 (7.89) & 4.46 \\
        & Offensive-Defensive Ratio ($X_2$) & 0.15 (0.21) & 0.14 (0.20) & 6.02 \\
        \midrule
        \multicolumn{5}{l}{\textit{Post-matching: ATT ($N = 45{,}046$ per group)}} \\
        \midrule
        & Distance to Endline ($X_1$) & 25.63 (7.83) & 25.91 (7.72) & 3.53 \\
        & Offensive-Defensive Ratio ($X_2$) & 0.15 (0.21) & 0.14 (0.21) & 4.27 \\
        \bottomrule
    \end{tabular}
\end{table}

\section{Additional real data analysis results}
\subsection{Propensity score model for the real data}
\begin{table}[h!]
    \small\sf\centering
    \caption{Propensity score model fitted by logistic regression.}
    \label{tab:ps_model}
    \begin{tabular}{lccc}
       \toprule
        & \textbf{Estimate} & \textbf{Std. Error} & \textbf{p-value} \\
        \midrule
        (Intercept) & -2.593 & 0.066 & $<$0.001 \\
        Score differential & -0.079 & 0.013 & $<$0.001 \\
        Ten minute warning & -0.063 & 0.034 & 0.061 \\
        Defender closing time & 0.005 & 0.006 & 0.375 \\
        Distance to sideline & 0.073 & 0.004 & $<$0.001 \\
        Teammate distance to goal line & -0.002 & 0.002 & 0.483 \\
        Distance to endline & -0.066 & 0.002 & $<$0.001 \\
        Offensive-defensive ratio in box & 2.042 & 0.064 & $<$0.001 \\
        Position: Midfielder & 0.505 & 0.041 & $<$0.001 \\
        Position: Defender & 1.184 & 0.045 & $<$0.001 \\
       \bottomrule
    \end{tabular}
\end{table}